\documentclass[twocolumn,showpacs,preprintnumbers,amsmath,amssymb,aps,prl]{revtex4}

\usepackage{graphicx}
\usepackage{bm}

\begin{document}

\title{Orbital-selective Mass Enhancements in Multi-band Ca$_{2-x}$Sr$_{x}$RuO$_{4}$
Systems Analyzed by the Extended Drude Model}

\author{J. S. Lee,$^{1}$ S. J. Moon,$^{1}$ T. W. Noh$^{1}$, S. Nakatsuji,$^{2}$ and
Y. Maeno$^{2,3}$}

\affiliation{$^{1}$ReCOE \& School of Physics, Seoul National
University, Seoul 151-747, Korea\\$^{2}$Department of Physics, Kyoto
University, Kyoto 606-8502,Japan\\$^{3}$International Innovation
Center, Kyoto University, Kyoto 606-8501, Japan}

\begin{abstract}
We investigated optical spectra of quasi-two-dimensional multi-band Ca$%
_{2-x} $Sr$_{x}$RuO$_{4}$ systems. The extended Drude model analysis
on the ab-plane optical conductivity spectra indicates that the
effective mass should be enhanced near $x=0.5$. Based on the sum
rule argument, we showed that the orbital-selective Mott-gap opening
for the $d_{yz/zx}$ bands, the widely investigated picture, could
not be the origin of the mass enhancement. We exploited the
multi-band effects in the extended Drude model analysis, and
demonstrated that the intriguing heavy mass state near $x=0.5$
should come from the renormalization of the $d_{xy}$ band.
\end{abstract}

 \maketitle

The Mott metal-insulator transition has been one of major topics in
condensed matter physics for many years. It is well known that the
conduction electron mass could be enhanced in metals due to strong
electron correlations, and that it should diverge toward a
metal-insulator transition. Through Brinkman and Rice, such a mass
divergence has been predicted to occur in bandwidth controlled
systems, where the ratio between the interaction strength $U$ and
the bandwidth $W$ can be varied for a fixed
band filling \cite{BR}. Recently, it has been shown that the isoelectronic Ca%
$_{2-x}$Sr$_{x}$RuO$_{4}$ (CSRO) compounds comprise an excellent
system to investigate the $W$-controlled Mott metal-insulator
transition \cite {Nakatsuji PRL,Friedt,HF214,Satoru PRL04}. Here $W$
for the Ru $t_{2g}$ states should keep decreasing due to the
internal chemical pressure from substituting the Sr$^{2+}$ ion with
the smaller Ca$^{2+}$ ion, which gives rise to a transition from a
multi-band metal ({\it x}=2.0) to a Mott-insulator ({\it x}=0.0).
Contrary to the simple expectation of the gradual change in $W$ with
{\it x}, these compounds exhibit quite unusual behavior near {\it
x}=0.5. While the metal-insulator transition occurs for {\it
x}$<$0.2, anomalous enhancements of magnetic susceptibility $\chi
(0)$ and specific heat coefficient $\gamma $, indicating a heavy
mass state, were observed near {\it x}=0.5 \cite {Nakatsuji
PRL,HF214}.

In order to explain these surprising experimental behaviors,
numerous workers have paid attention to the directional dependences
of the Ru $t_{2g}$ orbitals
\cite{Fang,Anisimov,Koga,Liebsch,ARPES05}. Due to the planar crystal
structure of the layered perovskite, partially filled $t_{2g}$ bands
become separated into a two-dimensional $d_{xy}$ band, and two
nearly one-dimensional $d_{yz/zx}$ bands. Since the $d_{xy}$ band at
\textit{x}=2.0 has a wider $W$ than the $d_{yz/zx}$ bands, some
workers have argued that orbital-selective Mott transition (OSMT)
could occur successively: i.e., at {\it x}=0.5 for the $d_{yz/zx}$
bands and at {\it x}$<$0.2 for the $d_{xy}$ band
\cite{Anisimov,Koga}. However, other workers insisted that the Mott
transitions should occur simultaneously for {\it x}$<$0.2
\cite{Liebsch}. The intriguing picture of OSMT has been the subject
of heated debate, but no consensus has been reached yet \cite
{ARPES05}.

Optical spectroscopy has been used as a powerful method to
investigate metal-insulator transitions of strongly correlated
electron systems. Especially, the extended Drude model (EDM) has
been extensively used to explain the electrodynamic responses near
metal-insulator transitions of numerous correlated electron systems,
such as heavy fermion compounds \cite {EDM} and high transition
temperature superconductors \cite{EDM2}. This
phenomenological theory describes optical conductivity spectra $\tilde{\sigma%
}(\omega )$ [=$\sigma _{1}(\omega )+i\sigma _{2}(\omega )$] of free carriers
in terms of the frequency($\omega $)-dependent scattering rate $1/\tau
(\omega )$ and the mass enhancement $\lambda (\omega )$ (effective mass $%
m^{\ast }(\omega )$ normalized by the band mass $m_{b}$), which is given as
\begin{equation}
\lambda (\omega )=\frac{\omega _{p}^{2}}{4\pi }\cdot \frac{\sigma
_{2}(\omega )}{\omega |\tilde{\sigma}(\omega )|^{2}}.
\end{equation}
Here, $\omega _{p}$ is the plasma frequency. However, it should be
noted that the free carriers in this model were assumed to originate
from a single band. Up to this point, there have been no efforts to
extend this analysis to the multi-band metallic compounds.

Even for a multi-band system, $\tilde{\sigma}(\omega )$ is an averaged
response for all $k$-space. It contains all the contributions from each
independent band, so it should be expressed as a linear addition of such
contributions. As a result, the optical sum rule should be valid for the
multi-band system, and the total spectral weight of the coherent peak $%
\omega _{p}^{2}$ should be $\omega _{p}^{2}=\sum_{i}\omega
_{p,i}^{2}$. By expressing the contribution from each band in terms
of EDM  and comparing the sum with Eq. (1), we obtained $\lambda
(\omega )=\sum_{i}\omega _{p,i}^{2}\sum_{i}\omega _{p,i}^{2}\tau
_{i}^{2}(\omega )\lambda _{i}(\omega )/\left( \sum_{i}\omega
_{p,i}^{2}\tau _{i}(\omega )\right) ^{2}$. When all of $\tau _{i}$
are nearly the same \cite{average}, $\lambda (\omega )$ obtained
from Eq.\ (1) corresponds approximately to the averaged value of
contributions from each band weighted by $\omega _{p,i}^{2} $ ,
i.e.,

\begin{equation}
\lambda (\omega )\simeq \sum_{i}\omega _{p,i}^{2}\lambda _{i}(\omega
)/\sum_{i}\omega _{p,i}^{2}.
\end{equation}
For the CSRO system, superscripts $\alpha $ and $\beta $ are commonly used
to indicate two bands formed by the $d_{yz}$ and the $d_{zx}$ orbital
states. And let us assign the $d_{yz/zx}^{\alpha }$, $%
d_{yz/zx}^{\beta }$, and $d_{xy}$ bands as $i$ = 1, 2, and 3, respectively.

In this Letter, we investigated the doping-dependent in-plane $\tilde{\sigma}%
(\omega )$ taking into account of the multi-band effects. Based on
the sum rule, we could argue that the controversial $d_{yz/zx}$
bands would not experience the OSMT. Instead, we could demonstrate
that the heavy mass state should be contributed to by a strong
renormalization of the $d_{xy}$ band. We will provide additional
indirect evidence for this intriguing picture from the
\textit{x}-dependent changes of $p-d$ charge transfer excitations.

High quality CSRO single crystals were grown by the floating zone method
\cite{Growth}. Near-normal incident reflectivity spectra were measured in a
photon energy range between 5 meV and 30 eV. Using the Kramers-Kronig
analysis, $\tilde{\sigma}(\omega )$ were obtained \cite{JSLEE}.

Figure 1 shows the doping-dependent ab-plane $\sigma _{1}(\omega )$
of CSRO (0.06$\leq ${\it x}$\leq $2.0) at room temperature. All of
the $\sigma _{1}(\omega )$ curves show strong coherent peaks in the
zero frequency limit. While the simple Drude model cannot explain
the $\omega $-dependence of these $\sigma _{1}(\omega )$,
application of the EDM would be useful to get information about the
changes of electrodynamic parameters.

\begin{figure}[tbp]
\includegraphics[width=2.4in]{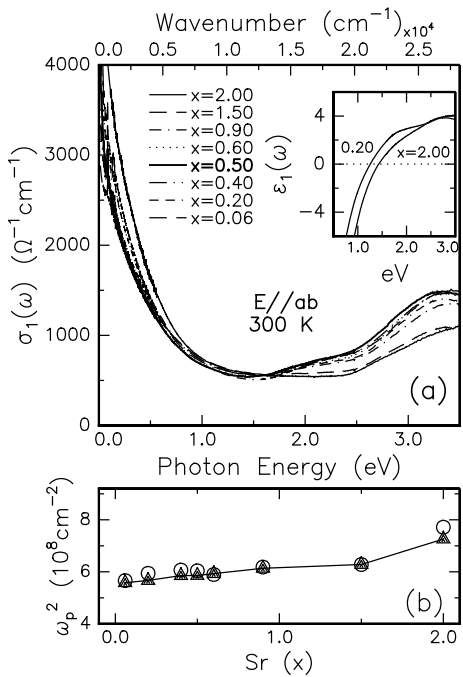}
\caption{(a) {\it x}-dependent $\protect\sigma _{1}(\protect\omega
)$ of Ca$_{2-x}$Sr$_{x}$RuO$_{4}$ at room temperature. Spectral
weight redistribution occurs around 1.5 eV. Inset shows $\protect\varepsilon _{1}(%
\protect\omega )$ of the {\it x}=2.00 and 0.20 compounds. (b) {\it
x} -dependent changes of $\protect\omega _{p}^{2}$. The solid and
open symbols represent values of $\protect\omega _{p}^{2}$
obtained
by the integration of $\protect\sigma _{1}(\protect\omega )$ and the zero-crossing of $\protect%
\varepsilon _{1}(\protect\omega )$, respectively.} \label{Fig.1}
\end{figure}

Before going into a detailed discussion about the mass enhancement
changes in CSRO using Eq.\ (1), it is worthwhile to mention how we
obtained $\omega _{p}$ values. First, we estimated $\omega _{p}$ by
integrating the spectral weight of the coherent peak using $\omega
_{p}^{2}=8\int_{0}^{\omega _{c}}\sigma _{1}(\omega )d\omega $. Here,
$\omega _{c}$ is the highest cutoff frequency for the free carrier
response. As shown in Fig.\ 1(a), an isosbetic point exists around
1.5 eV: as {\it x} decreases, the spectral weight below 1.5 eV
decreases, but that of the higher energy region increases. So, we
chose $\omega _{c}$ as 1.5 eV. The {\it x}-dependent values of
$\omega _{p}^{2}$ are displayed as solid triangles in Fig.\ 1(b),
where $\omega _{p}^{2}$ keeps decreasing down to {\it x}=0.06.
Second, we determined $\omega _{p}^{2}$ from $\varepsilon
_{1}(\omega )$ [=$\varepsilon_{\infty }-4\pi \sigma _{2}(\omega )/\omega $] using the relation $%
\varepsilon _{1}(\omega _{p}/\sqrt{\varepsilon _{\infty }})=0$.
$\varepsilon _{1}(\omega )$ for two representative metallic
compounds, i.e., {\it x}=2.00 and 0.20, are shown in the inset of Fig. 1(a). They show monotonic $%
\omega $-dependent increases crossing a zero level around 1.3 eV. By
adopting $\varepsilon _{\infty }=5.8$, we estimated $\omega _{p}^{2}$ and
displayed them as open circles in Fig.\ 1(b), which show behavior quite
similar to that of the solid triangles. For further discussion in this
paper, we will adopt the $\omega _{p}^{2}$ values estimated by the first
method.

\begin{figure}[tbp]
\includegraphics[width=2.4in]{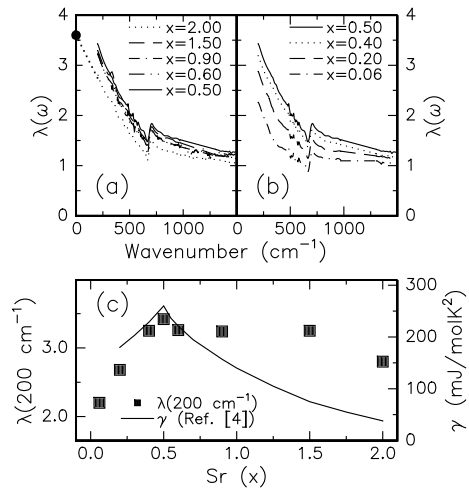}
\caption{$\protect\lambda (\protect\omega )$ of
Ca$_{2-x}$Sr$_{x}$RuO$_{4}$ at room temperature obtained by
extended Drude model analysis for (0.5$\leq ${\it x}$\leq $2.0)
(a). and for (0.06$\leq ${\it x}$\leq $0.5) (b). As {\it x}
decreases from 2.0, $\protect\lambda (\protect\omega )$ increases
initially, reaches a maximum at {\it x}=0.5, and then decreases.
(c) {\it x}-dependence of $\protect\lambda ($200 cm$^{-1})$ and
$\protect\gamma$, the latter of which is quoted from Ref.
\protect\cite {HF214}. Note that the enhancement of
$\protect\lambda ($200 cm$^{-1})$ at {\it x}=0.5 is similar to
that of $\protect\gamma$. However, its magnitude is much smaller.}
\label{Fig.2}
\end{figure}

Figures 2(a) and 2(b) show the results for $\lambda (\omega )$
obtained from Eq.\ (1). As {\it x} decreases, the $\lambda (\omega
)$ curves show a small upward shift down to {\it x}=0.5 [Fig.\
2(a)], and then they exhibit a\ rather significant downward shift
below {\it x}=0.5 [Fig.\ 2(b)]. Figure 2(c) shows the {\it
x}-dependent changes of $\lambda (\omega )$ at the low frequency
limit, i.e., 200 cm$^{-1}$, which displays a peak near {\it x}=0.5
\cite{10K}. This {\it x}-dependent change is reminiscent of the
enhancement of $\gamma $ shown as a solid line in Fig.\ 2(c) \cite
{HF214}. These unusual behaviors suggest that the Fermi surface
topologies must experience anomalies near {\it x}=0.5.

The de Haas-van Alphen (dHvA) measurement could be a very powerful
tool to investigate Fermi surface topologies of very pure samples.
Mackenzie {\it et al}. performed the dHvA measurements on
Sr$_{2}$RuO$_{4}$ \cite{Fermi liquid}. They observed quantum oscillations coming from three $%
t_{2g}$ bands, two from the $d_{yz/zx}$ bands and one from $d_{xy}$ band. They found that $n_{1}$:$n_{2}$:$n_{3}\simeq1$:$1$:$1$ and $m_{b,1}$:$%
m_{b,2}$:$m_{b,3}\simeq 1$:$2$:$3$, where $n_{i}$ is the free carrier
density of the $i$-th band. Since $\omega _{p}^{2}=4\pi ne^{2}/m_{b}$, $%
\omega _{p,1}^{2}$:$\omega _{p,2}^{2}$:$\omega _{p,3}^{2}$ $\simeq6$:$3$:$2$%
. And, they also reported that $\lambda _{1}(0)$:$\lambda
_{2}(0)$:$\lambda _{3}(0)\simeq 3.3$:$3$:$4$. From these values, we
can estimated $\lambda (0)$ as 3.3 from Eq.\ (2). This value is well
consistent with the optically observed value ($\sim $3.5), shown as
the solid circle on the y-axis of Fig.\ 2(a), demonstrating the
validity of our EDM analysis for the multi-band system.

On the other hand, angle-resolved photoemission spectroscopy (ARPES)
studies could be another way to look into the Fermi surface.
Recently, Wang {\it et al} performed ARPES measurements on an {\it
x}=0.5 sample, and showed that its Fermi surface topology remains
nearly the same as that of an {\it x}=2.0 sample \cite{ARPES05}.
They claimed that the OSMT should not occur in the CSRO system. The
photoemission studies are generally sensitive to the surface (not to
the bulk), and the ARPES study could not explain why the heavy mass
state does exist near the {\it x}=0.5 sample.

Optical spectroscopy could provide new insights on OSMT of the CSRO
system. We will test the model of the OSMT in the $d_{yz/zx}$ bands,
which is a seemingly fascinating and the most widely investigated
scenario \cite{Anisimov,Koga} to explain the anomalous heavy mass
state around {\it x}=0.5 \cite{HF214}. Using the dHvA measurement
results \cite{Fermi liquid}, the ratio between $\omega _{p,3}^{2}$
and $(\omega _{p,1}^{2}$+$\omega _{p,2}^{2})$ for {\it x}=2.0 could
be estimated to be about 1/4.5. If the $d_{yz/zx}$ bands open a Mott
gap, the optical sum rule states that the reduction in $\omega
_{p}^{2}$ should reach around 82 \% of $\omega _{p}^{2}$. However,
as shown in Fig.\ 1(b), $\omega _{p}^{2}$ for {\it x}=0.5 decreases
only by $\sim $20 \%. Such a difference in the reduction of $\omega
_{p}^{2}$ clearly demonstrates that the Mott-gap opening could not
occur in the narrow $d_{yz/zx}$ bands.

What should be responsible for the unusual behavior near {\it
x}=0.5? It should be noted that the enhancement of $\gamma $ between
the {\it x}=2.0 and the {\it x}=0.5 samples is about 6.0,
significantly higher than the corresponding value (i.e.\ 1.4) of
$\lambda ($200 cm$^{-1})$ [See Fig. 2(c).]. This difference comes
from how the mass enhancement of each band can contribute to the
total responses of $\gamma $ and $\lambda $ in the multi-band CSRO
system. The specific heat coefficient $\gamma $ can be written as a
linear summation of contributions from three $t_{2g}$ bands, i.e.,
$\gamma =\sum_{i}\gamma _{i}$. On the other hand, $\lambda (\omega
)$ obtained from Eq.\ (1) should be the averaged value of mass
enhancement contributions, weighted by $\omega _{p,i}^{2}$, as shown
in Eq.\ (2). As an illustration, let us consider a simple multi-band
system with three bands, whose plasma frequencies and masses are
exactly the same. If one of the bands experiences a mass enhancement
of 16 times, we can estimate that the total specific heat can be
enhanced by 6 times. And from Eq.\ (2), assuming $m_{b}=m^{\ast }$,
that the average $\lambda $ can be enhanced by only about 1.5 times.
In a real system, $\omega _{p}^{2}$ and $\lambda (\omega )$ of each
band cannot be the same. However, using a similar argument with
proper guidance, we will be able to unveil the nature of the heavy
mass state near the {\it x}=0.5 CSRO sample.

\begin{figure}[tbp]
\includegraphics[width=2.7in]{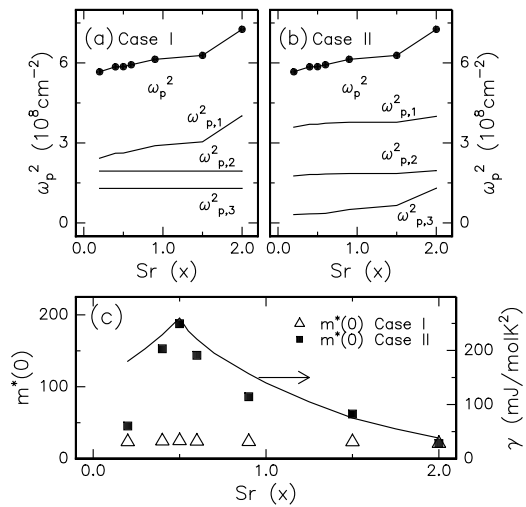}
\caption{(a) Proposed changes of $\protect\omega _{p}^{2}$ for
Case I, where only one of the $d_{yz/zx}$ bands experience a
reduction in spectral weight. (b) Proposed changes of of
$\protect\omega _{p}^{2}$ for Case II, where the $d_{xy}$ band
experiences a significant spectral weight change. (c) The
estimated $m^{\ast }(0)$ for each case. The solid line is the
specific heat data from Ref. \protect\cite{HF214}. The good
agreement between Case II and $\protect\gamma $ indicates that the
mass renormalization near the {\it x}=0.5 sample should occur mostly in the $%
d_{xy}$ band.} \label{Fig.3}
\end{figure}

First, let us assume one limiting case (Case I), where $\omega
_{p,1}^{2}$ would decrease with \textit{x,} while $\omega
_{p,2}^{2}$ and $\omega _{p,3}^{2}$ remain the same as those at {\it
x}=2.0. This limiting case is illustrated in Fig.\ 3(a). Using Eq.\
(2), $\omega _{p,i}^{2}=4\pi n_{i}e^{2}/m_{b,i}$ and $\lambda
_{i}(0)=m_{i}^{\ast }(0)/m_{b,i}$, we could evaluate $m_{i}^{\ast
}(0)$ \cite{comment-mass}, assuming the carrier densities $n_{i}$ of
each band would not change with {\it x} \cite {ARPES05}. As shown in
Fig.\ 3(c), $m^{\ast }(0) $ $[\equiv m_{1}^{\ast }(0)+m_{2}^{\ast
}(0)+m_{3}^{\ast }(0)]$ cannot be significantly enhanced, failing to
explain the specific heat data. The reduction of $\omega _{p,2}^{2}$
gives a similar result with that of Case I. Second, let us consider
the other limiting case (Case II), where the $d_{xy}$ band would
experience a significant change with {\it x} \cite{assume}, which is
illustrated in Fig.\ 3(b). Using a similar process, we could evaluate $%
m^{\ast }(0)$. As shown with the solid squares in Fig.\ 3(c),
$m^{\ast }(0)$ enhances from 21 at {\it x}=2.0 to around 200 at {\it
x}=0.5, and then decreases for {\it x}$<$0.5. This change of
$m^{\ast }(0)$ can explain that of $\gamma $ quite well. Our EDM
analysis based on the multi-band extension indicates that the heavy
mass state around {\it x}=0.5 should be contributed to by the
renormalization of the $d_{xy}$ band. It is quite interesting that the $d_{xy}$ band, having a larger $W$ than the $%
d_{yz/zx}$ bands, could be strongly renormalized near {\it x}=0.5.

\begin{figure}[tbp]
\includegraphics[width=2.4in]{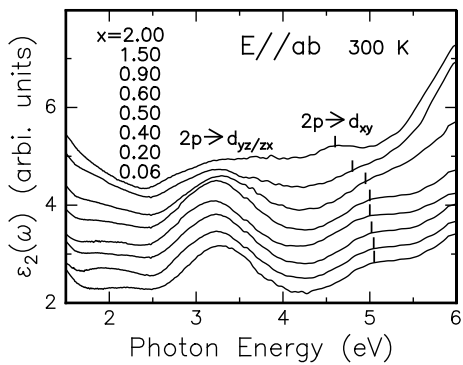}
\caption{\textit{x}-dependent ab-plane $\protect\varepsilon _{2}(\protect%
\omega )$ of the Ca$_{2-x}$Sr$_{x}$RuO$_{4}$ compounds. Two peaks
around 3.3
and 4.7 eV correspond to the transitions from the O 2$p$ band to the Ru $%
d_{yz/zx} $ and the Ru $d_{xy}$ band, respectively. The {\it
x}-dependent positions of the higher energy peak are indicated
with vertical solid lines. The shift of this O 2$p$ $\rightarrow$
Ru $d_{xy}$ peak indicates that the $d_{xy}$ band experience
significant electronic structural changes above {\it x}=0.5.}
\label{Fig.4}
\end{figure}

Finally, let us discuss how the $d_{xy}$ band could be renormalized
and give rise to a large $m^{\ast }$ near \textit{x}=0.5. Figure 4
shows {\it x}-dependent changes of $\varepsilon _{2}(\omega )$
[=$4\pi \sigma _{1}(\omega )/\omega $] from 1.5 eV to 6 eV. For the
\textit{x}=2.0 compound, there are multi-peak structures around 3.5
and 4.7 eV. For the metallic CSRO compounds, the RuO$_{6}$ octahedra
are largely elongated along the c-axis \cite{Friedt}. This leads to
a stronger hybridization of the $d_{xy}$ band with the O 2$p$ band
than that of the $d_{yz/zx}$ bands, and results in the
higher energy position of the $p-d$ transition to the $d_{xy}$ band \cite{pd}%
. So, the peaks around 3.5 and 4.7 eV can be attributed to the
charge-transfer transitions from the O 2$p$ band to the unoccupied Ru $%
d_{yz/zx}$ bands and to the Ru $d_{xy}$ band, respectively. As {\it
x} decreases down to {\it x}=0.5, the lower energy peak does not
experience a discernible change. However, the higher energy peak
shows a strong blue shift, indicating that the unoccupied band
having the $d_{xy}$ character must be shifted to a higher energy
away from the Fermi energy. These changes could result in smaller
contribution of $d_{xy}$ band to the metallic
response and lead to the enhancement of its $m^{\ast }(0)$ near \textit{x}%
=0.5. From the neutron scattering experiments \cite{Friedt}, it is
well known that a rotation of the RuO$_{6}$ octahedra along the
c-axis increases from around \textit{x}=1.5 and saturates below
\textit{x}=0.5. In this respect, the observed evolution of the
$d_{xy}$ band could be attributed to the RuO$_{6}$ rotation
\cite{origin}. The detailed mechanism should be elucidated in future
work \cite{origin2}.

In summary, we addressed the evolution of electronic structures in
multi-band Ca$_{2-x}$Sr$_{x}$RuO$_{4}$ by investigating the optical
spectra. The argument based on the optical sum rule revealed that
$d_{yz/zx}$ bands should remain itinerant near \textit{x}=0.5. Using
the extended Drude model analysis, we found that the Fermi surface
topologies should experience anomalies near {\it x}=0.5, and we
proposed that the heavy mass state near {\it x}=0.5 should come from
a strong renormalization of the $d_{xy}$ band.

We acknowledge the Pohang Advanced Light Source for allowing us to use some
of their facilities. This work was supported by the Ministry of Science and
Technology through the Creative Research Initiative program and in part by
Grants-in-Aid for Scientific Research from JSPS of Japan.


\begin{references}
\bibitem{10K}  We obtained the optical spectra at 10 K for the {\it x}=2.0,
0.5, and 0.2 compounds, and found that the behaviors at 10 K are similar
with those at room temperature.
\bibitem{comment-mass}  $\lambda (0)$ for each {\it x} is linearly
extrapolated from the value at $200$ cm$^{-1}$ as for the case {\it
x}=2.0 shown in Fig. 2(a). And, {\it x}-dependent $m_{1}^{\ast
}(0)$, $m_{2}^{\ast }(0)$, and $m_{3}^{\ast }(0)$ are normalized as
they should be to 3, 6, and 12, respectively, at {\it x}=2.0
\cite{Fermi liquid}.
\bibitem{assume}  If we assumed that the spectral change with {\it x}
might be solely attributed to the $d_{xy}$ band, we found negative values of $%
\omega _{p}^{2}$ below {\it x}=0.9. So we considered that the
$d_{yz/zx}$ bands also have small changes, while the main spectral
changes should come from the change of the $d_{xy}$ band.
\bibitem{origin}  Figure 3(c) shows that the estimated $m^{\ast }(0)$ values
in Case II agree quite well with the experimental $\gamma $ values
for the samples with {\it x}$\geq$0.5. However, there are some
deviations for the samples with {\it x}$<$0.5. This deviation could
be attributed to the tilting of the RuO$_{6}$ octahedra, which would
provide additional increase of $\gamma $ \cite{Friedt}.
\end{references}
\end{document}